\documentclass[12pt, a4paper]{article}

\usepackage[utf8]{inputenc}
\usepackage{geometry}
\usepackage{amsmath, amssymb, amsfonts}
\usepackage{graphicx}
\usepackage{float}
\usepackage{booktabs}
\usepackage{hyperref}
\usepackage{caption}
\usepackage{subcaption}
\usepackage{authblk}
\usepackage{setspace}

\title{\textbf{Metaboplasticity: The Reciprocal Regulation of Neuronal Activity and Cellular Energetics}}

\author[1]{Ece Öner}
\author[2]{Cenk Denktaş}

\affil[1]{Department of Molecular Biology and Genetics, Yildiz Technical University, Istanbul, Turkey}
\affil[2]{Department of Physics, Yildiz Technical University, Istanbul, Turkey}

\date{} 

\begin{document}

\maketitle

\vspace{-1em}

\noindent\textbf{RESEARCH ARTICLE}

\vspace{1em}

\noindent\textbf{Correspondence:}  
Cenk Denktaş (\href{mailto:cdenktas@yildiz.edu.tr}{cdenktas@yildiz.edu.tr})

\vspace{0.5em}

\noindent\textbf{Funding:}  
The authors received no specific funding for this work.

\vspace{0.5em}

\noindent\textbf{Keywords:}  
Neuroenergetics \;STDP \; Metaboplasticity \;$Q_{10}$ Scaling \; STDP Kernel Deformation 

\vspace{2em}

\begin{abstract}
Due to the complexities of simulating bioenergetics, most Spiking Neural Network (SNN) models treat neurones as energetically unconstrained, omitting the metabolic costs of neural activity. We introduce a computational framework that employs thermodynamic variables to directly integrate metabolic regulation into network dynamics. Using temperature-dependent Q10 scaling as a biophysical proxy for ATP availability, we developed a conductance-based leaky integrate-and-fire microcircuit (5,000 neurones) in Brian2 in which membrane leak conductance, and therefore the membrane time constant, emerges from temperature-dependent kinetics. This strategy translated metabolic state into intrinsic excitability across a physiologically relevant thermal sweep without requiring exhaustive biochemical pathways. The model simulated non-linear metabolic modulation of Hebbian plasticity, such as hypermetabolic fluctuations resulting in seizure-like avalanche dynamics and hypometabolic inhibition of LTP. The results of this study demonstrate that differential Q10 scaling provides a biologically valid and computationally efficient approach for incorporating metabolic variables into large-scale neuronal simulations.
\end{abstract}

\section{Introduction}

The human brain’s metabolic needs to support its computational functions are considerably greater than the requirements of other organs as compared to its size\textsuperscript{1}. More than 20\% of the human body’s total oxygen and glucose are consumed in the brain, thereby making the brain the most energetically demanding organ despite making up only a small fraction of total body mass\textsuperscript{4}.

This substantial metabolic expenditure is not static. The biophysical mechanisms of neuronal signalling and information processing cause most of the great demand for energy. Up to 75\% of this energy in the cortical grey matter is utilised by the $\text{Na}^+/\text{K}^+$-ATPase pump, supplying electrical signalling in order to maintain the resting membrane potential\textsuperscript{1}. The initiation and propagation of action potentials\textsuperscript{7}, the steps of synaptic transmission, such as neurotransmitter synthesis, vesicle recycling, and postsynaptic receptor trafficking, are the most energetically demanding processes in the brain\textsuperscript{2}.

This rapid metabolic cycle necessitates an elaborate, concurrent connection among neural activity, cerebral blood flow, and metabolic supply.\textsuperscript{4} Glial cells, particularly astrocytes, regulate the mentioned neurovascular connection by functioning as an essential bridge between the vasculature and synapses. Astrocytes transport energy substrates such as lactate and glucose to neurones in response to neuronal activity, thus guaranteeing the sensitive metabolic supply and demand balance\textsuperscript{1}.

Mitochondria contributes largely to the subcellular regulation of neuronal energetics. This regulation is not static; mitochondria are mobile organelles which localise in metabolically demanding regions such as presynaptic terminals and postsynaptic dendritic spines\textsuperscript{14}. The localisation of mitochondria is critical for rapidly supplying the ATP required for processes like synaptic vesicle recycling\textsuperscript{17} and for supporting the local protein synthesis required for prolonged structural modifications associated with synaptic plasticity.\textsuperscript{15} Furthermore, localised synaptic mitochondria are key regulators of transient $\text{Ca}^{2+}$ elevations.

Transient $\text{Ca}^{2+}$ elevations are crucial both for neurotransmitter release and for the induction of plasticity mechanisms, including long-term potentiation (LTP) and long-term depression (LTD).\textsuperscript{15} This organisation of neuronal energy dynamics introduces an intrinsic dependency between synaptic activity and local metabolic strength. High-frequency stimulation and the accompanying $\text{Ca}^{2+}$ influx\textsuperscript{15} initiate the cellular mechanisms of learning and building memories, namely LTP\textsuperscript{20}, are processes which demand high levels of energy. The generation of high-frequency spike trains\textsuperscript{9} and the buffering of large calcium loads\textsuperscript{15} are ATP-expensive processes, supplied by localised synaptic mitochondria\textsuperscript{14}. Therefore, the probability of the Hebbian plasticity depends on the metabolic condition of the synapse. Even under ideal Hebbian conditions, if metabolic resources are scarce, plasticity cannot occur.

\subsection{Metabolic Regulation of Inherent Neuron Excitability}
Neuronal excitability is an essential biophysical parameter of the availability of the energy within the system. The certain classes of ion channels and ion pumps, which are allosterically regulated by the intracellular ATP/ADP ratio of energy change, build this connection. This process is mediated by two main molecular sensors, namely $\text{Na}^+/\text{K}^+$-ATPase and the K-ATP channel.

$\text{Na}^+/\text{K}^+$-ATPase is the primary consumer of cellular ATP\textsuperscript{5} as mentioned before, and thus its function is directly dependent on the availability of ATP. Under low-metabolic conditions, such as in hypoxia, a decrease in the energy levels can lead to a malfunction of the ion pumps. This malfunction causes an increasing loss of the $\text{Na}^+$ and $\text{K}^+$ gradients, and a catastrophic depolarisation of the resting membrane potential (RMP).\textsuperscript{24}

ATP-sensitive potassium (K-ATP) channels are more acute.\textsuperscript{27} These channels express the energy state within the cell more directly.\textsuperscript{28} When the cells have high ATP levels, K-ATP channels are allosterically inhibited, thus remaining closed. However, during periods of metabolic stress (e.g., hypoxia or glucose deprivation), the decrease in the ATP levels removes this inhibition, allowing channels to open\textsuperscript{29}. The opening of K-ATP channels pushes the RMP away from the action potential threshold and closer to the Nernst equilibrium potential for potassium ions. This is called a state of hyperpolarisation. The hyperpolarised neurone is less excitable and requires a much stronger synaptic input signal to depolarise the membrane to the firing threshold.\textsuperscript{30} By silencing the action potential generation, the cell reduces its primary energy expenditure, the $\text{Na}^+/\text{K}^+$ pump, and conserves the limited ATP reserves.

This mechanism provides a direct, biophysical link between metabolic state and the intrinsic excitability of a neurone.\textsuperscript{32} In vivo, the energy-excitability coupling is observed in pathological and extreme physiological states. The two extremes, namely hypometabolism and hypermetabolism, are dissected in this study.

Hypometabolism is the state of low energy seen in conditions such as hypoxia and hypothermia. Acute hypometabolism behaves as a protective brake on neural excitability, as established before.\textsuperscript{24} On the other hand, chronic hypoxia has severe pathological consequences. Constant metabolic stress can lead to permanent cognitive deficits and irregular network development by disrupting the molecular mechanisms of synaptic plasticity.\textsuperscript{34} As suggested from hypothermia, this state can be conceptualised through the biophysical effects of low temperature.

Hypermetabolism is the state of high energy, occurring in fevers and seizures. The hypermetabolic state can be conceptualised through the biophysical effects of high temperature. For most biological reactions, the $Q_{10}$ temperature coefficient describes the temperature dependence. $Q_{10}$ temperature coefficient is usually defined as the rate ratio of a reaction increasing for every 10°C rise in temperature.\textsuperscript{38} For ion channel kinetics and synaptic processes, $Q_{10}$ values are typically between 2 and 3\textsuperscript{40}, meaning that even a small increase in the brain temperature can cause unstable, significant acceleration in the neural process rates.

A state of hypermetabolism can lead the network into a state of pathological hyperexcitability\textsuperscript{43}, most commonly observed as febrile seizures in the developing brain.\textsuperscript{44} Though it may be naively assumed that more neuronal activity would lead to more plasticity, the opposite is proven true. The febrile seizures in animal models demonstrate that the hyperexcitable state causes damage in LTP and leads to the formation of irregular synaptic circuits\textsuperscript{45}. Both extreme models suggest an ``inverted-U'' relationship between metabolic rate and plasticity. The hypometabolic state would be expected to flatten the plasticity curve\textsuperscript{31}. The normometabolic state stands as an optimal point for Hebbian learning, where the neuronal firing is informative and structured. The hypermetabolic state represents a pathological escape condition. The high, non-informative firing rates would likely push for an explosive, non-specific potentiation and result in a dysfunctional state of maximum-strength, all-to-all connectivity that lacks the sparse, refined computational structure of a healthy, informed network, a state consistent with the damaged plasticity seen in vivo.\textsuperscript{45} Thus, both metabolic extremes are detrimental to functional network development.

\subsection{Unifying Model of Metaboplasticity}
The synapse-specific Hebbian rules, such as STDP, and homeostatic mechanisms behind the stabilisation of network-wide activity has been studied extensively\textsuperscript{22}. Spike-Timing-Dependent Plasticity (STDP) is the biological implementation of the Hebbian rule and is the core of learning in Spiking Neural Networks (SNNs)\textsuperscript{70}. In the STDP model, the timing of spikes is crucial. If a presynaptic (sender) neuron fires a spike before the post synaptic (receiver) neuron fires, the connection between the two neurons is strengthened. This process is named Long-Term Potentiation, or LTP for short. Conversely, if the postsynaptic neuron fires just before the presynaptic one, the connection is weakened (Long-Term Depression, or LTD).\textsuperscript{50} This mechanism provides an unsupervised learning algorithm which allows a network to learn from the inputs\textsuperscript{69}.

STDP rules can be formulated in various different ways when it comes to the computational models. One of the ways to formulate STDP is to introduce additive rules, where the weight change is fixed regardless of the current synaptic weight. Another way is achieved by using multiplicative rules, where the change depends on the current weight, providing more stability.\textsuperscript{87} These rules form the computational basis for how networks are assumed to learn and store information.\textsuperscript{50}

Metabolic state is also recognised as a regulator. For example, AMP-activated protein kinase (AMPK) is recognised as negative regulator of late-phase LTP (L-LTP).\textsuperscript{52} When cellular energy is low, AMPK activates and inhibits the mTOR pathway, thus suppressing the signaling cascade essential for the protein synthesis of building long-term memory.\textsuperscript{55} Similarly, mitochondrial dysfunction is known to be the primary driver of synaptic pathology and damaged plasticity in numerous neurodegenerative disorders, including Alzheimer's and Parkinson's disease.\textsuperscript{17}

However, a computational framework that unifies the metabolic constraints with network dynamics remains elusive.\textsuperscript{64} We suggest that integrating metabolic influences into Hebbian learning principles does not require designing a new plasticity rule. The metabolic perturbations could directly adjust neuronal intrinsic excitability. Changes in excitability modify the network’s baseline firing rate, leading a standard Hebbian STDP rule without alteration. The network’s emergent connectivity can reshaped with by altering the frequency of pre- and post-synaptic spike pairings.

Our study tests this hypothesis in silico by implementing a spiking neural network (SNN). In our model, the intrinsic excitability parameters of the neurons are modulated to simulate metabolic state. The metabolic states are implemented as temperature. Hypometabolism is achieved by 293.15 Kelvin (20°C), hypermetabolism is achieved by 307.15 Kelvin (34°C) meanwhile the control group, normometabolism is achieved by 300.15 Kelvin (27°C). The other parameters are held constant.

\subsection{Limitations of the Metabolic Proxy}
In this study, temperature is used as a biophysically grounded proxy for metabolic state rather than a direct model of cellular ATP dynamics. Temperature directly modulates ion channel kinetics, synaptic time constants, and plasticity rates via Q10 scaling, whereas metabolic state influences neuronal function primarily through ATP availability and downstream signaling pathways. While these processes are tightly coupled in vivo, they are not identical. Accordingly, the present model captures the kinetic consequences of metabolic constraints on excitability and plasticity, rather than explicitly simulating biochemical energy metabolism. All interpretations are therefore restricted to the dynamical and thermodynamic consequences of altered metabolic capacity.

\section{Methodology}

The complete in-silico simulation and the following analyses were completed using the Brian 2 software, version 2.6.01. Brian 2 is an open-source simulation software for spiking neural networks. The simulations were configured to use the `numpy' computational target for repeatability, with a simulation time step of $0.5 \, \text{ms}$. The network architecture and the neuronal models were designed to approximate canonical cortical microcircuits.\textsuperscript{13}

\subsection{Spiking Neural Network (SNN) Model}
\subsubsection{Network Architecture and Topology}
The simulated network consisted of $N=5000$ neurons. The neurons were split into two non-overlapping classes: 4000 excitatory neurons and 1000 inhibitory neurons. This classification was in accordance with Dale’s principle\textsuperscript{16}. The 4:1 ratio of excitatory to inhibitory neurons in particular is a standard biological constraint in cortical microcircuit models\textsuperscript{18} and is requisite for balanced dynamic systems.

A random connectivity pattern was generated connecting neurons within and between the two aforementioned classes. With a uniform random connection probability of $p=0.02$, four distinct synaptic pathways were defined:
\begin{itemize}
    \item Excitatory-to-Excitatory (E-to-E)
    \item Excitatory-to-Inhibitory (E-to-I)
    \item Inhibitory-to-Excitatory (I-to-E)
    \item Inhibitory-to-Inhibitory (I-to-I)
\end{itemize}
Except for E-to-E connections, the remaining connections are non-plastic (static) and maintain the network's dynamic Excitatory-Inhibitory balance. The weights for the static connections are defined as fixed conductances: $w_{ei,static}= 4.0 \, \text{nS}$, $w_{ie,static}= 12.0 \, \text{nS}$, $w_{ii,static}= 5.0 \, \text{nS}$.

\subsubsection{Neuronal Dynamics: The Conductance-Based Leaky Integrate-and-Fire (gLIF) Model}
The individual neurons in the simulation were modelled as conductance-based Leaky Integrate-and-Fire (gLIF) neurons.\textsuperscript{6} This model was selected for the purpose of achieving greater biophysical realism as in gLIF model, synaptic inputs modulate specific membrane conductances ($g_e$ and $g_i$), the resulting synaptic current is voltage-dependent.

\begin{equation}
    I_{syn} = g_e(E_e - V) + g_i(E_i - V) \label{eq:1}
\end{equation}

The subthreshold dynamics of the membrane potential $V$ for each neuron are controlled by the given differential equation in equation two, as defined in our algorithm:
\begin{equation}
    C_m \frac{dV}{dt} = g_{l,sim}(E_l - V) + g_e(E_e - V) + g_i(E_i - V) \label{eq:2}
\end{equation}

In equation two, $g_e$ and $g_i$ are the total instantaneous excitatory and inhibitory conductances, defined as part of the neuron model, allowing them to be summed and integrated at each time step as they evolve based on synaptic inputs.\textsuperscript{21} The term $g_{l,sim}$ is the leak conductance derived from baseline $g_l$, scaled by temperature. The other constant biophysical parameters are listed in Table 1.

\begin{table}[H]
\centering
\caption{Neuron Model Parameters}
\begin{tabular}{@{}llc@{}}
\toprule
\textbf{Parameter} & \textbf{Description} & \textbf{Value} \\ \midrule
$C_m$ & Membrane capacitance & 200 pF \\
$g_l$ & Baseline leak conductance (at $T_{ref}$) & 10.0 nS \\
$E_l$ & Leak reversal potential (rest potential) & -60 mV \\
$E_e$ & Excitatory reversal potential & 0 mV \\
$E_i$ & Inhibitory reversal potential & -80 mV \\
$V_{th}$ & Firing threshold & -50 mV \\
$V_r$ & Reset potential & -60 mV \\
$\tau_{refractory}$ & Refractory period & 5 ms \\ \bottomrule
\end{tabular}
\end{table}

The neuron fires an action potential when the membrane potential ($V$) crosses the defined firing threshold of -50 mV ($V_{th}$). Two instantaneous events occur following the spike: the membrane potential resets to a fixed value of -60 mV, the neuron enters an absolute refractory period.

In gLIF, $\tau_m$ is not an independent parameter but is an emergent property\textsuperscript{22} defined by the following equation:
\begin{equation}
    \tau_m = \frac{C_m}{g_l} \label{eq:3}
\end{equation}

An important feature of the neuron model in this study is the relationship between the leak conductance ($g_l$) and intrinsic membrane time constant ($\tau_m$). At baseline, the equation yields $\tau_m = 200 \, \text{pF} / 10 \, \text{nS} = 20 \, \text{ms}$. The leak conductance is known to be sensitive to temperature\textsuperscript{26}, this will be detailed in Section II of the methodology. Therefore, temperature modulation is applied directly to $g_l$, and $\tau_m$ is altered as a mechanistic consequence. As temperature increases, ion channels become more active, $g_l$ increases, and $\tau_m$ decreases, aligning with the natural model\textsuperscript{5}.

\subsubsection{Synaptic Dynamics and External Drive}
The dynamics of the summed excitatory and inhibitory conductances in the neuron model are governed by the arrival of presynaptic spikes. Between spike events the conductances show an exponential decrease towards zero.\textsuperscript{6} Conductance values increment instantaneously upon the arrival of presynaptic spikes.

An external population of 1000 neurons firing at 5 Hz is implemented to maintain the network in an asynchronous firing regime. These external inputs project onto both excitatory neuron and inhibitory neurons with a connection probability of 0.1 and fixed conductances of 5.0 nS and 2.5 nS respectively.

\subsection{Modelling Temperature Dependency Via $Q_{10}$ Scaling}
In this study, we established three distinct metabolic states of the human brain (hypermetabolic, normometabolic, hypoxic) by applying different temperature conditions. Temperature condition is implemented using the $Q_{10}$ temperature coefficient, a unitless metric which is standard for quantifying the temperature dependence of a rate process.\textsuperscript{36}

The $Q_{10}$ coefficient is defined as the factor by which the rate of a biological process $R$ increases for a 10°C or 10 K rise in the temperature $T$.\textsuperscript{36} The scaling function is defined for any given parameter based on its $Q_{10}$ value and defined reference temperature $T_{ref}$:
\begin{equation}
    \Phi(T) = Q_{10}^{\frac{T - T_{ref}}{10^\circ C}} \label{eq:4}
\end{equation}

The updated, temperature-dependent value of the parameter $X(T)$ is then calculated at every temperature $T$ with the following equation:
\begin{equation}
    X(T) = X_{ref} \cdot \Phi(T) \label{eq:5}
\end{equation}

This notation has been applied to all biophysical parameters in the rate-based and temperature-sensitive models. In our model, we implemented two distinct $Q_{10}$ factors and a reference temperature of 300.15K. One of the $Q_{10}$ factors in the model scales membrane and synaptic time constants by 2.0. The other $Q_{10}$ factor scales the amplitudes of the STDP learning rule by 1.5.

\subsubsection{Temperature Effects on Neuronal Excitability}
Temperature modulation in neurons essentially targets the kinetics of the ion channels.\textsuperscript{40} This shows as a modulation of membrane conductances, including the leak conductance $g_l$\textsuperscript{26}.
\begin{equation}
    g_{l,sim}(T) = g_{l,ref} \cdot Q_{10,m}^{\frac{T - T_{ref}}{10^\circ C}} \label{eq:6}
\end{equation}

As established in section 3.1.2 the membrane time constant is dependent on the leak conductance. The membrane capacitance is assumed to be temperature-invariant. This leads to an emerging scaling relationship for the membrane time constant:
\begin{equation}
    \tau_m(T) = \frac{C_m}{g_{l,sim}(T)} = \tau_{m,ref} \cdot Q_{10,m}^{-\frac{T - T_{ref}}{10^\circ C}} \label{eq:7}
\end{equation}

\subsubsection{Temperature Effects on Synaptic Kinetics}
The kinetics involving in the closing of synaptic ion channels are also sensitive to the changes in the temperature. The mentioned kinetics determine the decay time constants of excitatory and inhibitory inputs. At higher temperatures, neurons have shorter membrane time constants and integrate shorter synaptic inputs, increasing the rapidity of the network regime.
\begin{equation}
    \tau_e(T) = \tau_{e,ref} \cdot Q_{10,m}^{-\frac{T - T_{ref}}{10^\circ C}} \label{eq:8}
\end{equation}
\begin{equation}
    \tau_i(T) = \tau_{i,ref} \cdot Q_{10,m}^{-\frac{T - T_{ref}}{10^\circ C}} \label{eq:9}
\end{equation}

\subsection{Temperature-Dependent Model of STDP}
This temperature-dependent spike-timing-dependent plasticity mechanism is implemented exclusively on the recurrent Excitatory-to-Excitatory synapses. The STDP model is based on a standard, pair-based, additive STDP rule\textsuperscript{35}, modified to make its parameters subject to temperature regulation.

The synaptic weights updates based on the relative timing of pre-synaptic and post-synaptic spikes. The spikes are tracked using two variables that are local to each synapse, notated as $A_{pre}$ and $A_{post}$\textsuperscript{7}. The traces are modelled as exponentially decayed variables.
\begin{equation}
    \frac{dA_{pre}}{dt} = -\frac{A_{pre}}{\tau_{pre,sim}} \label{eq:10}
\end{equation}
\begin{equation}
    \frac{dA_{post}}{dt} = -\frac{A_{post}}{\tau_{post,sim}} \label{eq:11}
\end{equation}

The baseline time constants for pre-synaptic and post-synaptic spikes are defined as equal, 20 ms. This defines the width of the STDP window. The time constants are scaled by kinetic factor $Q_{10,m} = 2.0$.
\begin{equation}
    \tau_{pre}(T) = \tau_{pre,ref} \cdot Q_{10,m}^{-\frac{T - T_{ref}}{10^\circ C}} \label{eq:12}
\end{equation}
\begin{equation}
    \tau_{post}(T) = \tau_{post,ref} \cdot Q_{10,m}^{-\frac{T - T_{ref}}{10^\circ C}} \label{eq:13}
\end{equation}

\subsubsection{Synaptic Weight Update Rule and Metabolic Scaling}
The synaptic weight is normalised between 0 and 1.0, modulates maximum conductance on both pre-synaptic and post-synaptic spikes. Synaptic weights are updated using a standard pair-based STDP rule with hard bounds. Conductance and trace variables are updated instantaneously upon spike arrival. The amplitudes of the STDP learning rate are scaled with $Q_{10}$ coefficient of 1.5. The symmetric scaling models metabolic regulation of the plasticity amplitudes\textsuperscript{46}.

\subsection{In-Silico Experimental Protocol and Analysis}
All simulations are run for a duration of 30 seconds, with the ``euler'' numerical integration method. Network states are randomised at the beginning of each simulation run. Membrane potentials are initialised to a random values near the resting potential of 5 mV. All plastic Excitatory-to-Excitatory synaptic weights are initialised to a random value drawn from a uniform distribution.

The simulation is executed across a three-dimensional parameter sweep involving 60 unique simulations to critically disentangle kinetic and plastic effects. This experiment represents three different metabolic states with three distinct thermal regimes:
\begin{itemize}
    \item \textbf{Hypometabolic/Hypoxic}: 293.15K
    \item \textbf{Control/Normometabolic}: 300.15K
    \item \textbf{Hypermetabolic}: 307.15K
\end{itemize}

These metabolic states are tested against two plasticity conditions (STDP-ON, STDP-OFF) across ten unique random seeds per condition. ``STDP-ON'' condition simulates full biophysical model, combining both kinetic effects of temperature on channel conductances and time constants with the metabolic scaling of the STDP rule. In contrast, ``STDP-OFF'' condition is designed to be the control condition. The latter condition is designed to isolate the pure effect of temperature on network kinetics and quantify its impact on excitability, independent of plastic rewiring.\textsuperscript{10}

\section{Results}

\subsection{Implementation of Metabolic Constraints}
We implemented a computational simulation model where synaptic kinetics are strictly modulated by thermodynamic laws to establish a biophysical foundation for understanding metabolism dependent synaptic plasticity. Figure 1 illustrates the conceptual scheme wherein the plasticity window is dynamically regulated by temperature ($T$). The model applies a $Q_{10}$ scaling factor 2.0 to the time constants and 1.5 to the learning rate amplitudes, as discussed in the methodology section in this paper.
The equations in Figure 1 depict that the time constant scales inversely with the temperature, implying that hypermetabolic states result in more rapid synaptic decays. On the other hand, the amplitude of the learning rate scales directly, modelling the increased effectiveness of neurotransmitter release and receptor trafficking at higher temperatures. The formulation embeds metabolic state directly into the learning rule as a fundamental biophysical constraint, instead of relegating it to a secondary homeostatic mechanism.

\begin{figure}[H]
    \centering
    \includegraphics[width=0.8\textwidth]{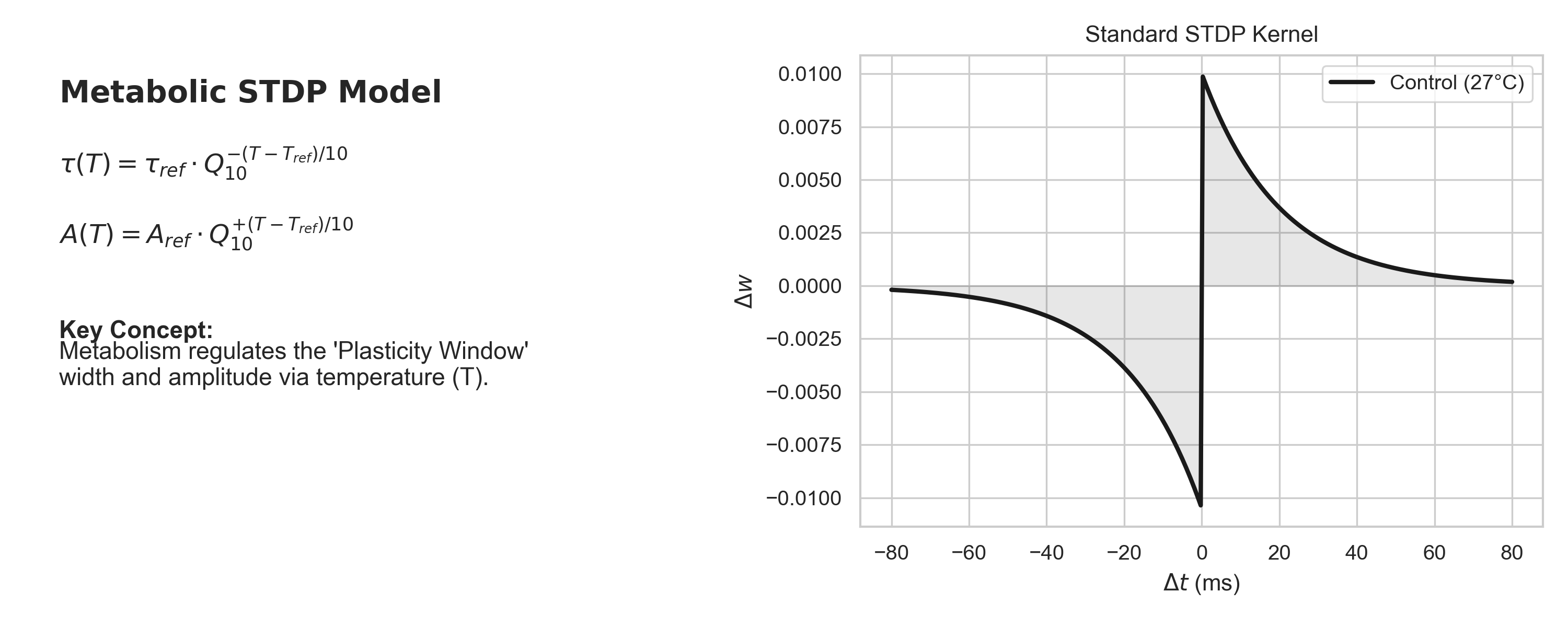} 
    \caption{Schematic outline of the computational implementation of temperature-dependent Spike-Timing-Dependent Plasticity (STDP). The left panel presents the modulating equations where the synaptic time constant $\tau(T)$ and learning rate amplitude $A(T)$ are modulated by temperature (T) via a Q10 scaling factor relative to a reference temperature of 27°C. The right panel displays the asymmetric STDP kernel under the reference temperature, defined over a temporal window of $\Delta t \in [-80, 80]$ ms. The kernel shows a casual LTP branch for positive $\Delta t$ values ($\Delta t>0$) with a peak amplitude of $\Delta w = 0.01$. For negative $\Delta t$ values an anti-casual LTD branch occurs ($\Delta t<0$), with a peak depression of $\Delta w= -0.01$.}
    \label{fig:1}
\end{figure}

\subsection{Emerging Metabolic Trade-Off Landscape}
By analysing the relationship between metabolic expenditure, defined as population firing rate, and learning success, defined as synaptic weight potentiation, the network-level consequences of the established kinetic constraints were evaluated. Figure 2 reveals a bifurcation in network dynamics.
The Cost-Function correlation on Figure 2A shows that excitability regimes are discrete. Hypometabolic simulations (purple) are strictly confined to a low-frequency domain around 4.75Hz and 5.00Hz, therefore yielding minimal weight changes around 0.202. In contrast, hypermetabolic simulations (orange) cluster in a high-cost, high-frequency domain ($>6.25$Hz), with a significant increase in mean synaptic weight around 0.215.

The plasticity evolution in Figure 2 tracks the temporal divergence of hypometabolic and hypermetabolic states over the 30-seconds simulation period. The control condition (red) maintains a stable, linear learning trajectory. However the hypermetabolic state (orange) exhibits a non-linear acceleration curve, convexly deviating from linearity at $t=5$s, indicating a positive feedback loop where increased connectivity facilitates increased firing rate which in return drive further potentiation. Conversely, the hypometabolic trace (purple) plateaus near baseline values, confirming that energy scarcity imposes a thermodynamic ceiling on synaptic growth.

\begin{figure}[H]
    \centering
    \includegraphics[width=\textwidth]{Fig2_Correlation_Traces.png} 
    \caption{Delineating the metabolic cost-benefit landscape governing synaptic plasticity. (A) Cost-Function Correlation: Positive, non-linear correlation between metabolic expenditure and synaptic efficacy. (B) Plasticity Evolution: Temporal divergence of hypometabolic and hypermetabolic states over the simulation epoch. Hypometabolic simulations yield minimal weight changes, while hypermetabolic simulations cluster in high-cost, high-frequency domains.}
    \label{fig:2}
\end{figure}

\subsection{Single-Cell Sensitivity}
The spike-timing-dependent plasticity kernel was analysed at the single-synapse level to explain the biophysical mechanism underneath the macroscopic instability we encountered in Figure 2. Figure 3 quantifies the deformation of the learning window under thermal stress.
The hypermetabolic trace (orange) exhibits a significant sharpening of the plasticity window. The decay time constant is reduced by approximately 1.6, relative to the baseline, consistent with $Q_{10}=2.0$ over a 7°C shift, while the peak amplitude is increased. The result is high-gain regime with a narrow temporal window, yielding highly precise but strong coincidence detection that destabilizes synaptic weights.
Under hypometabolic conditions (purple), the trace is flattened and prolonged. The reduce in amplitude results in a low-gain state that suppresses stochastic spike pairings, and acts as a homeostatic inhibitory process, preventing potentiation when metabolic resources are limited.

\begin{figure}[H]
    \centering
    \includegraphics[width=0.8\textwidth]{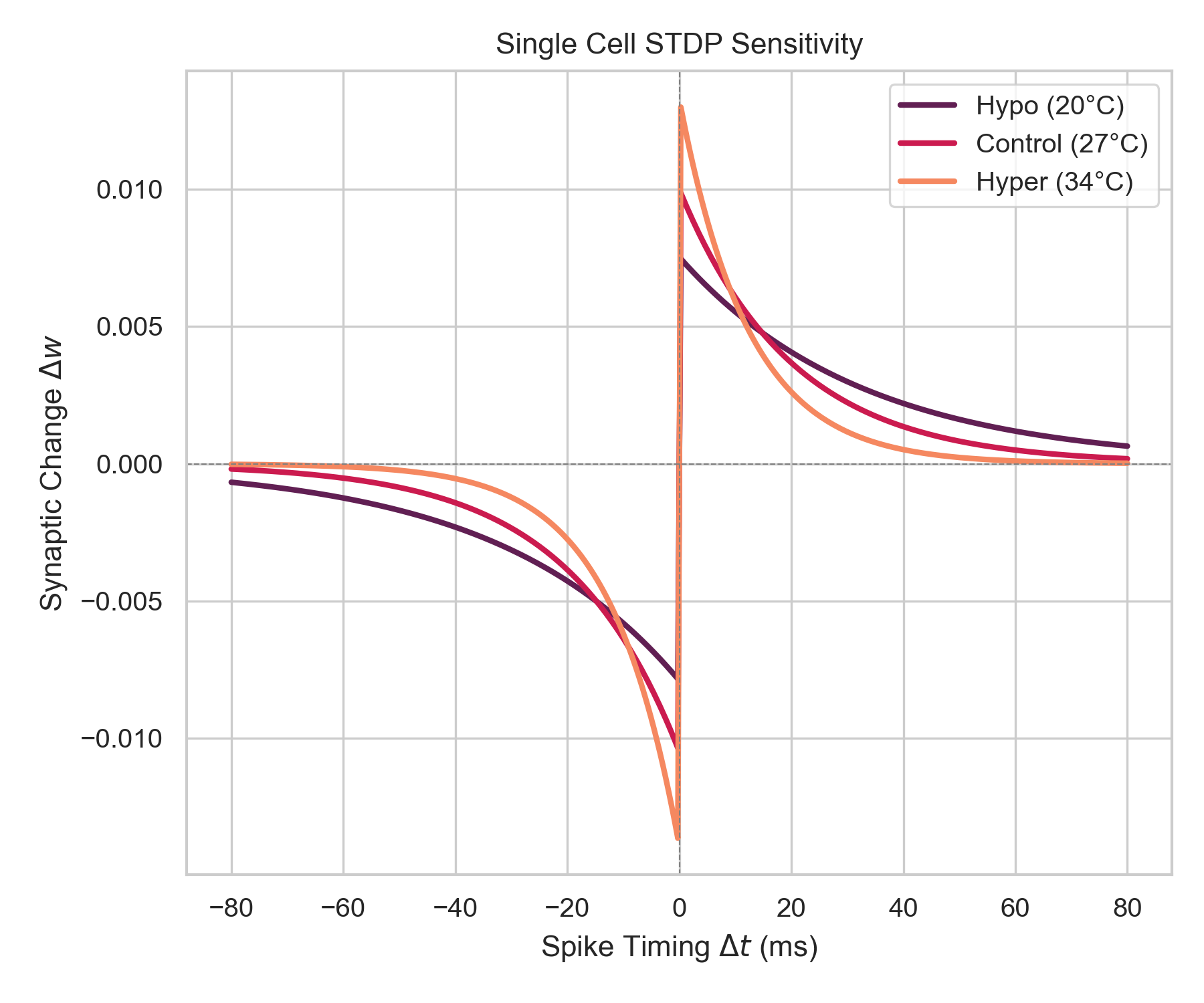} 
    \caption{Figure 3 illustrates the explicit deformation of the STDP learning rule under the three metabolic states. The y-axis ($\Delta w$) spans between -0.015 to 0.015, and x-axis covers [-80, 80] milliseconds time interval. The hyper kernel (orange) shows a sharpened profile with a peak amplitude increasing to around 0.013 and a reduced time constant ($\tau < 20$ms). Hypo kernel shown in purple displays a flattened profile with peak amplitude attenuated to around 0.007 and a broadened temporal window ($\tau > 20$ms), suggesting increased temporal integration but reduced synaptic efficacy per spike pair. The control kernel retains the standard parameters defined in Figure 1.}
    \label{fig:3}
\end{figure}

\subsection{Metabolic Scaling and Signal Degradation}
The relationship between metabolic intensity and final structural connectivity was further quantified in Figure 4, depicting the mean synaptic weight as a proxy for recall accuracy across the three simulated thermal regimes.
The data indicates a strictly linear positive correlation between metabolic state and connection strength, with mean synaptic weights increasing from hypometabolic 0.202 to control state of 0.205 to hypermetabolic 0.215. While the variance across the groups seem to be low, the steep ascent in the hypermetabolic condition implies a saturation of the weight matrix. Although the magnitude of connectivity scales with metabolism, this result should be interpreted in the context of network dynamics. Such widespread potentiation suggest a deterioration of sparse-coding specificity, shifting the network from a structure optimised for storing information into one of pathological over-connectivity.

\begin{figure}[H]
    \centering
    \includegraphics[width=0.8\textwidth]{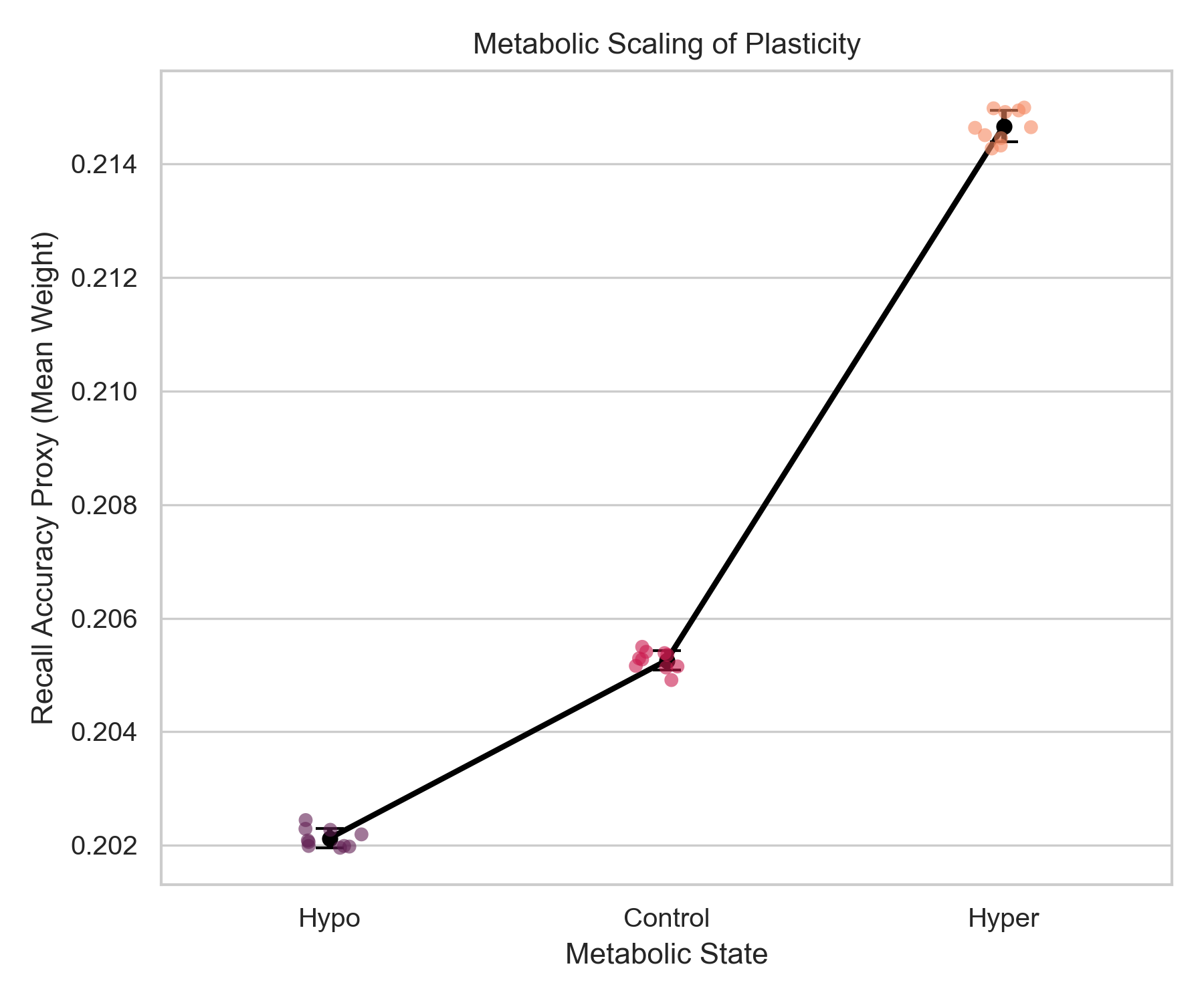}
    \caption{Figure 4 uses mean synaptic weight as a proxy for recall accuracy and summarises the statistical relationship between metabolic state and final network connectivity. Data points represent individual simulation trials (N=10 per condition). The hypometabolic condition yields the lowest mean weights, slightly above 0.202 while the control condition shows intermediate potentiation around 0.205. The hypermetabolic state condition achieves maximum synaptic weight above 0.214. Error bars denote the standard deviation across trials.}
    \label{fig:4}
\end{figure}

\subsection{Divergence of Firing Regimes and Topology}
The signal degradation hypothesis is confirmed by the spatiotemporal analysis of spiking activity, presented in Figure 5B, 5C and 5D. The raster plots dissect microscopic probability distributions of synaptic weights and macroscopic firing patterns of the network.

Figure A reveals a fundamental reshaping of the connectivity landscape, led by metabolic constrains. The hypometabolic distribution is leptokurtic, exhibits a negligible right-hand tail. This shape implies a systemic failure in differentiation, where synaptic weights remain clustered around the initialisation mean due to the dampened STDP rule. The absence of potentiation results in a homogenous network, lacking the structural variance required for memory formation.

Conversely, the hypermetabolic distribution (orange) exhibits a heavy rightward skew with a pronounced tail, extending toward the upper bound ($w=1.0$). While this confirms the successful potentiation, the sheer breadth of the distribution indicates a loss of selectivity. An increased number of synapses have been potentiated indiscriminately, suggesting that the accelerated kinetics have supressed the competitive nature of Hebbian learning, learning to a saturation of the synaptic weight matrix.

In Figure 5B, Hypo raster graph displays distinct horizontal striations. This topology indicates that while a subset of neurons maintain atonic activity, likely driven by external input, this excitation fails to propagate laterally within the recurrent network. The hyperpolarised resting potential raises the threshold for activation, functionally isolating active neurons and preventing the cell-assembly formation.

In sharp contrast, the Hyper raster graph in Figure 5D exhibits strong vertical banding. The vertical alignments are characteristic of large-scale synchronisation events, in which activity generates almost instantaneously. The shift from horizontal isolation to vertical synchrony visually demonstrates the collapse of the network into a pathological, hyper-synchronous attractor. In this state, the individual coding capacity of neurons is overridden by global oscillatory dynamics, effectively erasing the network’s meaningful information content.

\begin{figure}[H]
    \centering
    \includegraphics[width=\textwidth]{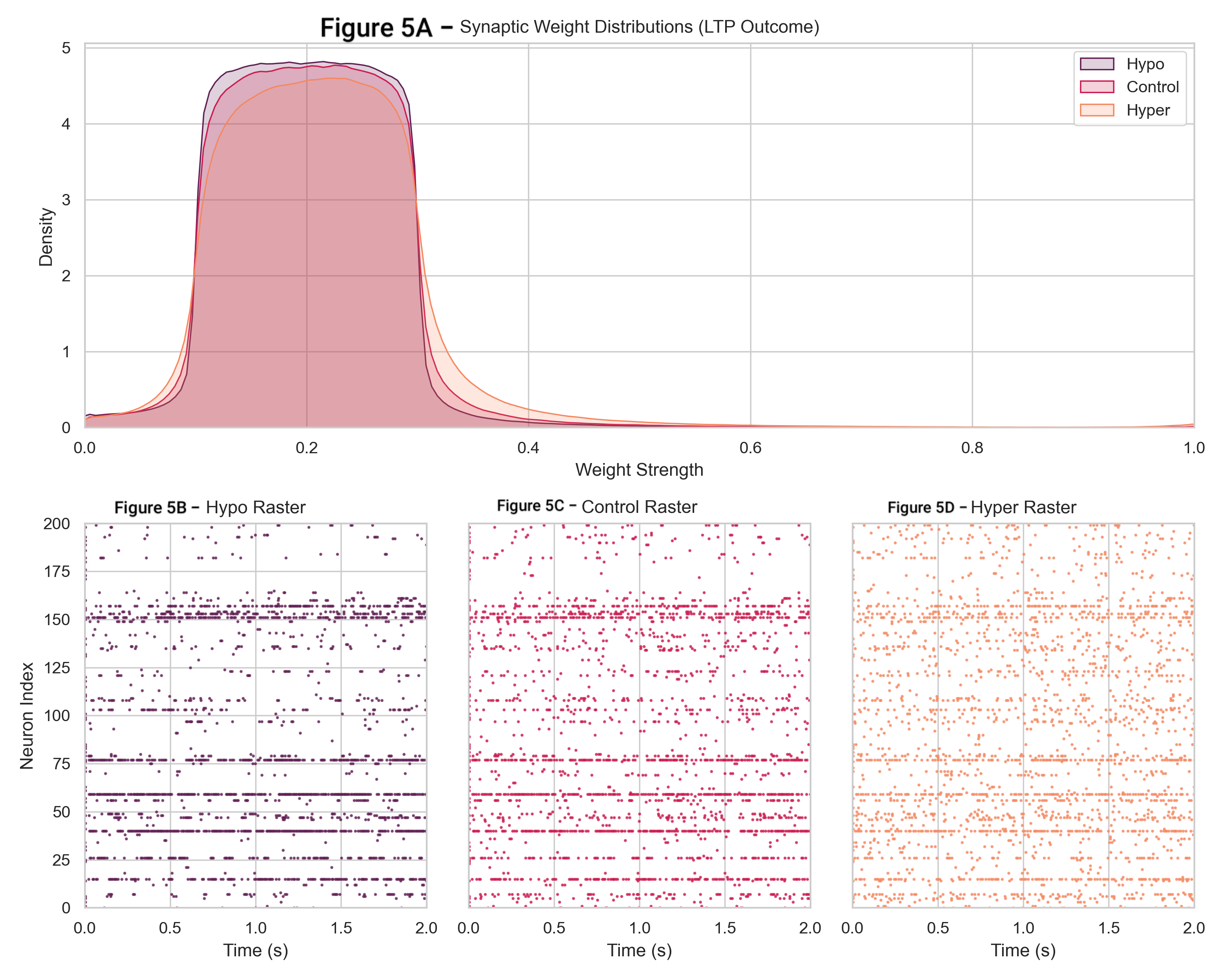} 
    \caption{(A) Probability density functions of synaptic weights post-simulation. The distribution of hypometabolic (purple) is leptokurtic and right-skewed, peaking around synaptic weight strength of 2.0 with a heavy tail truncation around w=4.0, indicating failed potentiation. The hypermetabolic (orange) is platycurtic with a significant mass shifted towards w>0.3, reflecting successful LTP. (B-D) Raster plots for a subset of neurons ($n=200$) over a 2 second interval. The hypo raster in Figure 5B shows sparse, asynchronous firing while the hyper raster in Figure 5D reveals dense, synchronous bursting.}
    \label{figure:5}
\end{figure}

\subsection{Parametric Robustness}
Sensitivity analysis was confirmed to ensure the divergent regimes were consequences of the metabolic state, rather than results of random network initialisation. The final mean weights across ten distinct random seeds (0-9) is shown on Figure 6. The resulting heatmap displays horizontal banding. For any given state, the variance in the final synaptic weight is negligible ($\Delta w<0.001$), whereas variance between metabolic states is statistically significant, therefore confirming that the network’s attractor state is a robust, inherent property of the biophysical parameters, independent of the specific initial micro-connectivity.

\begin{figure}[H]
    \centering
    \includegraphics[width=0.8\textwidth]{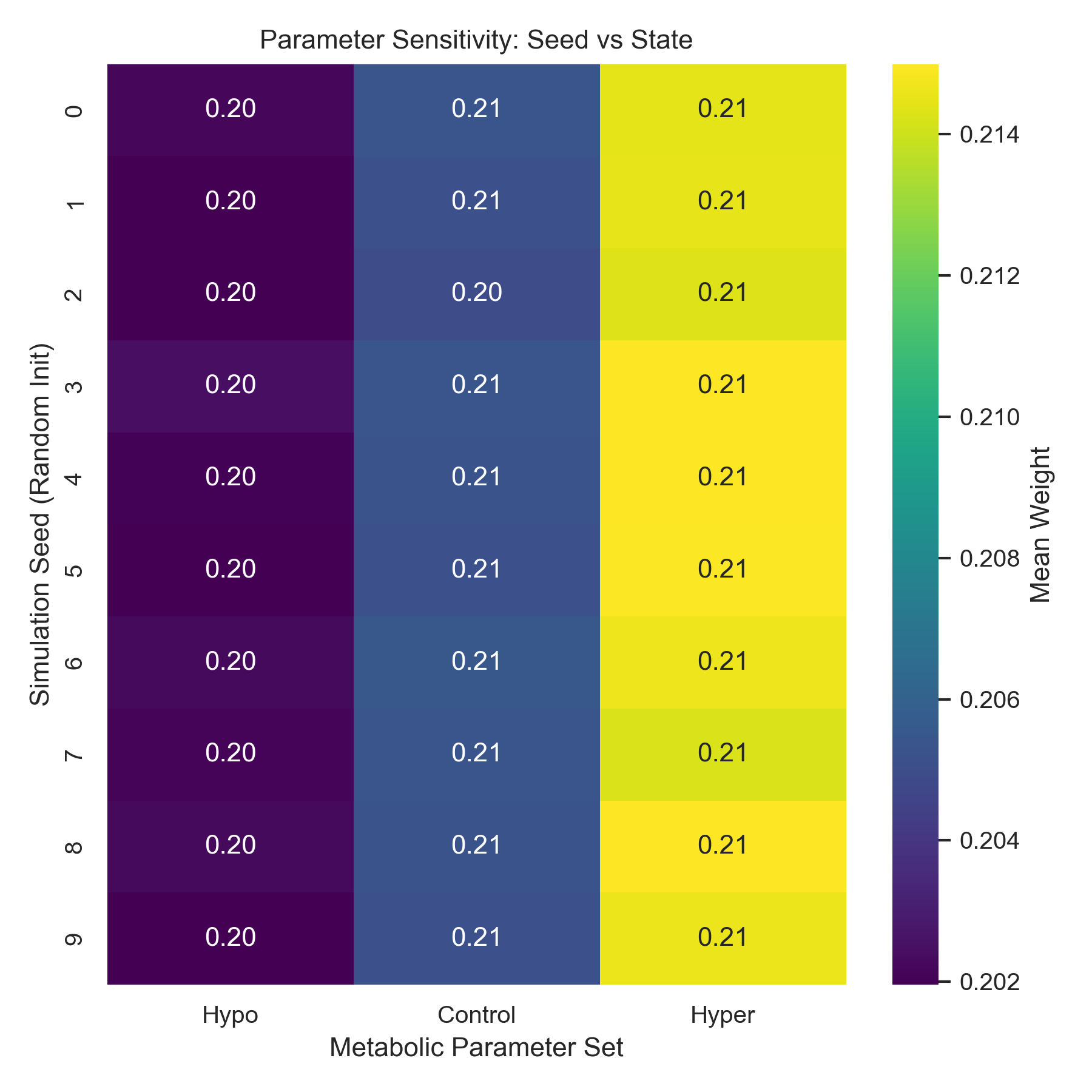} 
    \caption{Parameter sensitivity analysis: Heatmap of mean synaptic weight as a function of the Metabolic Parameter Set and stochastic initialization seeds. The strictly columnar stratification of the data demonstrates zero variance across simulation seeds.}
    \label{figure:6}
\end{figure}

\section{Discussion}

\subsection{The ``Inverted-U'' of Functional Plasticity}
The results discussed in the previous section provide mechanistic evidence for the ``Inverted-U" relationship between metabolic rate and neuroplasticity, extending the concept to include information theoretic constraints. While Figure 4 indicates a linear increase in synaptic connectivity with temperature, the integration of network dynamics shown in Figure 5 suggests that functional utility is maximised only at normometabolism. The hypermetabolic state achieves maximum connectivity with synaptic mean weight around 0.215, however this occurs with a collapse in information entropy. The hypersynchrony observed in the hypermetabolic raster plot in Figure 5D suggests that the conditional probability of any neuron firing given the firing of its neighbours approaches unification. In such state, the network’s capacity of representing diverse information states is minimised, as all units convey redundant information. This finding challenges the assumption that enhanced plasticity metrics in isolation equate to improved cognitive function. The optimal state for Hebbian learning is a confined thermal range where excitability is sufficient to permit plasticity but regulated enough to preserve the sparse coding necessary for information storage.

\subsection{Integration of Cellular Energetics and Network Dynamics}
A central contribution of this study is the establishment and empirical validation of a unifying framework in which metabolic constraints operate as fundamental hardware regulators for the software dynamics of Hebbian learning. In this model, intrinsic excitability is mechanistically coupled to the timing of the spike-timing-dependent plasticity (STDP) window through Q10 kinetics, demonstrating that the brain’s ability to encode information is fundamentally dependent on its thermodynamic and energetic state. Under hypometabolic conditions, hyperpolarisation and the flattening of the integration window prioritise energy conservation over synaptic potentiation, cellular and network survival over the formation of new memory links. This mechanism provides a biophysically grounded explanation for the regulation of the substantial ATP costs associated with information storage, which are estimated to operate at 4.75-5.5 Hz within the stable model. Importantly, these results indicate that metaboplasticity (metabolism dependent plasticity) is not only a regulatory factor or a secondary adaptation of the learning rule, rather an essential requirement for Hebbian plasticity to occur. Additionally, this framework demonstrates how metabolic state intrinsically shapes the temporal dynamics of learning. The STDP window is dynamically scaled according to metabolic conditions, meaning that the efficiency, timing precision, and amplitude of synaptic modifications are directly gated by the energy available to the neuron. This integration of energetic constraints with synaptic plasticity provides a mechanistic lens for understanding the emergent trade-offs between information storage, network stability, and energy homeostasis.

\subsection{Intrinsic Stability Compared to Externally Enforced Homeostasis}
The stability observed in the normometabolic group calls attention to the efficacy of biological constraints relative to externally enforced controls. Many spiking neural network (SNN) models require explicit, extrinsic homeostatic normalisation terms to prevent runaway potentiation. In our model, stability emerges fundamentally from the thermodynamics of the ion channels. The interplay between the refractory period, the membrane time constant and STDP window, all scaled by temperature, creates a natural modulator on network activity. The stable linear growth trajectory observed in the control condition contrasts with the exponential acceleration of the hypermetabolic state, demonstrating that network homeostasis is an emergent property of thermodynamic equilibrium. This implies that biological systems may rely on “hardware" limitations, such as energetics, over “software" limitations such as algorithmic normalisations, to maintain stability.

\subsection{Pathophysiological Implications and Clinical Correlations}
The validity of the model is supported by its alignment with known clinical phenomena. The synaptic suppression seen in hypometabolic simulations mirrors the mechanisms underlying therapeutic hypothermia. Clinically, hypothermia is induced to reduce excitotoxic damage after cardiac arrest. Our results suggest that this neuroprotective effect arises from the thermodynamic slowing of synaptic kinetics and the flattening of the STDP window, which together prevent the spread of excitotoxic cascades and the formation of maladaptive synaptic patterns.

In contrast, the model also captures the mechanistic basis of febrile seizures. Elevated core temperatures act not merely as systemic stressors but as direct modulators of neuronal excitability, shortening refractory periods via the Q10 effect and driving the network into the global synchronization observed in Figure 5. Additionally, the metabolic gating of plasticity revealed here offers a mechanistic explanation for cognitive deficits associated with chronic hypoxia. Under metabolic constraints, the biophysical “hardware” effectively freezes the synaptic weight matrix to conserve energy, thereby restricting the encoding of new information and limiting learning capacity.

\section{Conclusion}

This study establishes that metaboplasticity, the reciprocal regulation of neuronal activity and cellular energetics, is not merely a homeostatic background process. It is a fundamental determinant of neural computation. By using temperature-dependent Q10 kinetics as a biophysically grounded proxy for metaolic constrains, we demonstrated that excitability and Hebbian learning are mechanistically constrained by the bioenergetic capacity of the cell. Our simulations revealed a critical "energy-excitability manifold" where hypometabolic states imposed a protective silence by flattening the plasticity window. In contrast, hypermetabolic states drove the network into a collapse of representational diversity, characterised by high connectivity but with highly redundant population activity, effectively mimicking seizure dynamics.

Two key implications follow from these findings. First, they demonstrate that network stability can emerge spontaneously from thermodynamic hardware constraints rather than needing artificial software algorithms for normalization. This offers a more biologically plausible model of homeostasis. Second, they highlight that maximum synaptic potentiation does not equate to maximum function. Instead, optimal learning memory formation occur only within a narrow thermodynamic window where excitability, plasticity, and energy expenditure are balanced.

\section{Future Outlook}

Looking forward, this framework unlocks a several promising avenues for both neuroscience and artificial intelligence research. In clinical neuroscience, metabolism-aware neural modelling may provide a powerful tool for understanding disorders in which energetic dysregulation precedes overt structural damage, such as epilepsy, neurodegenerative diseases, and mitochondrial channelopathies. If combined with neuroimaging techniques which probe metabolic and functional states, such as fMRI, PET or MR spectroscopy, AI-based models incorporating metabolic constrains could enable the early detection of pathological network regimes. If applied at the onset of disease, such approaches may provide prognostically valuable information by identifying maladaptive dynamical trajectories before irreversible network reorganisation occurs. In parallel, the results have direct implications for neuromorphic engineering and biologically inspired AI. Incorporating virtual metabolic constraints into learning architectures may offer a principled solution to persistent challenges such as runaway plasticity, catastrophic forgetting, and energy inefficiency in artificial neural systems. By embedding energetic costs directly into the learning dynamics, future AI systems may achieve greater stability, efficiency, and robustness, more closely mirroring the operating principles of biological brains. Ultimately, by bridging the gap between cellular bioenergetics and network topology, this work lays the foundation for a new class of neural models that account for the true physical cost of a thought.

\end{document}